\documentclass{aastex63}

\usepackage{amsmath}
\usepackage{gensymb}
\graphicspath{{./}{figures/}}
\usepackage{graphicx}
\usepackage{multirow}

\begin{document}

\title{No Detectable Kilonova Counterpart is Expected for O3 Neutron Star--Black Hole Candidates}

\author[0000-0002-9195-4904]{Jin-Ping Zhu}
\affil{Department of Astronomy, School of Physics, Peking University, Beijing 100871, China}

\author[0000-0002-9188-5435]{Shichao Wu}
\affiliation{Department of Astronomy, Beijing Normal University, Beijing 100875, China}

\author[0000-0001-6374-8313]{Yuan-Pei Yang}
\affiliation{South-Western Institute for Astronomy Research, Yunnan University, Kunming, Yunnan, People’s Republic of China}

\author[0000-0002-9725-2524]{Bing Zhang}
\affiliation{Department of Physics and Astronomy, University of Nevada, Las Vegas, NV 89154, USA}

\author[0000-0002-1067-1911]{Yun-Wei Yu}
\affiliation{Institute of Astrophysics, Central China Normal University, Wuhan 430079, China}

\author[0000-0002-3100-6558]{He Gao}
\affiliation{Department of Astronomy, Beijing Normal University, Beijing 100875, China}

\author[0000-0002-1932-7295]{Zhoujian Cao}
\affiliation{Department of Astronomy, Beijing Normal University, Beijing 100875, China}

\author[0000-0002-8708-0597]{Liang-Duan Liu}
\affiliation{Department of Astronomy, Beijing Normal University, Beijing 100875, China}

\correspondingauthor{Jin-Ping Zhu; Bing Zhang; Shichao Wu; Zhoujian Cao} 
\email{zhujp@pku.edu.cn; zhang@physics.unlv.edu; wushichao@mail.bnu.edu.cn; zjcao@amt.ac.cn}

\begin{abstract}
We analyse the tidal disruption probability of potential neutron star--black hole (NSBH) merger gravitational wave (GW) events, including GW$190426\_152155$, GW190814, GW$200105\_162426$ and GW$200115\_042309$, detected during the third observing run of the LIGO/Virgo Collaboration, and the detectability of kilonova emission in connection with these events. The posterior distributions of GW190814 and GW$200105\_162426$ show that they must be plunging events and hence no kilonova signal is expected from these events. With the stiffest NS equation of state allowed by the constraint of GW170817 taken into account, the probability that GW$190426\_152155$ and GW$200115\_042309$ can make tidal disruption is $\sim24\%$ and $\sim3\%$, respectively. However, the predicted kilonova brightness is too faint to be detected for present follow-up search campaigns, which explains the lack of electromagnetic (EM) counterpart detection after triggers of these GW events. Based on the best constrained population synthesis simulation results, we find that disrupted events account for only $\lesssim20\%$ of cosmological NSBH mergers since most of the primary BHs could have low spins. The associated kilonovae for those disrupted events are still difficult to be discovered by LSST after GW triggers in the future, because of their low brightness and larger distances. For future GW-triggered multi-messenger observations, potential short-duration gamma-ray bursts and afterglows are more probable EM counterparts of NSBH GW events.
\end{abstract}


\section{Introduction} \label{sec:intro}

Mergers of binary neutron star (BNS) and neutron star--black hole (NSBH) have been proposed to be progenitors of short-duration gamma-ray bursts \citep[sGRBs;][]{eichler1989,paczynski1991,narayan1992}, source of heavy elements \citep{lattimer1974,lattimer1976} and kilonovae \citep{li1998,metzger2010}. On 2017 August 17th, a gravitational wave (GW) signal from a binary neutron star (BNS) merger event \citep[GW170817;][]{abbott2017GW170817} was first identified by the GW detectors of LIGO and Virgo. Subsequent data analyses revealed that this BNS GW event was associated with a sGRB \citep[GW170817A;][]{abbott2017gravitational,goldstein2017,savchenko2017,zhang2018}, a fast-evolving ultraviolet-optical-infrared transient \citep[AT2017gfo; e.g.,][]{abbott2017multi,arcavi2017,coulter2017,drout2017,evans2017,kasliwal2017,pian2017,smartt2017,kilpatrick2017}, and a broadband off-axis jet afterglow \citep{margutti2017,troja2017,lazzati2018,lyman2018,ghirlanda2019}. The fast-evolving transient was well explained by the theoretical predictions of kilonova emission \citep{li1998,metzger2010}. The joint observations of the GW signal of this BNS event and its associated electromagnetic (EM) counterparts provided the smoking-gun evidence for the long-hypothesized origin of sGRBs and kilonovae. 

The multi-messenger observations of GW170817 provided an ideal paragon for GW-led follow-up searches of EM signals. With the successful detection of a GW signal and its associated EM counterparts from a BNS merger, one may especially expect to catch as-yet undiscovered GW signals from neutron star--black hole (NSBH) mergers and search for their associated kilonova emissions in upcoming observing runs for the LIGO/Virgo collaboration (LVC). During the third observing run (O3) of LVC, several GW alerts for compact binary coalescence (CBC) with at least one NS member in the merger system (i.e., either BNS or NSBH mergers) have been issued. However, in spite of many efforts for follow-up observations, no confirmed EM counterpart candidate was identified\footnote{\cite{goldstein2019} reported a sub-threshold GRB candidate, GBM-190816, which was potentially associated with a sub-threshold LVC compact binary coalescence candidate. \cite{yang2020} investigated the physical implications of this potential association and indicated this GW candidate could be a NSBH merger with the mass ratio $Q = 2.26^{+2.75}_{-1.43}$. \cite{li2021} showed this event could have a large effective spin. However, since both the GW candidate and sGRB candidate were sub-threshold, the case was not confirmed.} \citep[e.g.,][]{anand2020,andreoni2020,coughlin2020b,gompertz2020,page2020,kasliwal2020,sagues2020,becerra2021}. \cite{kawaguchi2020b,fragione2021} gave a constraint on the NS mass-radius relationship based on the observed results of no confirmed counterpart for NSBH candidates. One plausible explanation for the lack of detection of an EM counterpart is that present EM searches are too shallow to achieve distance and volumetric coverage for the probability maps of LVC events \citep{coughlin2020a,sagues2020,zhu2020b}. However, it is also possible and even likely that the EM counterparts are intrinsically missing, e.g. for NSBH mergers there might not be material left outside of the merged BH remnant since the NS is not tidally disrupted but is swallowed as a whole into the BH (i.e. the so-called ``plunging'' event) \citep{zhu2020b}. 

Very recently, the observations of four CBC GW events in O3 of LVC with component masses potentially consistent with NSBH binaries have been reported. These four potential NSBH events are GW$190426\_152155$ \citep{abbott2020gwtc2}, GW190814\footnote{GW190814's secondary component of mass lies in the range of $2.50-2.67\,M_\odot$ which is at a high confidence level above the maximum allowed mass of a NS \citep{abbott2020search}, but the possibility of a NS secondary component cannot be excluded \citep[e.g.,][]{godzieba2020,zhang2020}. Therefore, we still include this event in our discussion.} \citep{abbott2020GW190814}, GW$200105\_162426$, and GW$200115\_042309$ \citep{abbott2021NSBH}, abbreviated as GW190426, GW190814, GW200105, and GW200115 hereafter. Many detailed follow-up observations for these GW events show no possible associated EM counterpart \citep[e.g.,][]{hosseinzadeh2019,goldstein2019b,anand2020,thakur2020,kilpatrick2021,alexander2021}. In this paper, we analyse the tidal disruption probability of these four NSBH GW events to investigate why their associated EM signals were not detected by the follow-up observations. Moreover, by combining population synthesis results, we assess the detectability of NSBH EM counterparts for future GW-triggered follow-up observations.

\section{Tidal Disruption and Kilonova Detectability}

\subsection{Fitting Formulas for the Remnant Mass and Dynamical Ejecta Mass}
Whether a NS can be tidally disrupted by the primary BH and eject a certain amount of material could be determined by the relative positions between the radius of the BH innermost stable circular orbit (ISCO) $R_{\rm ISCO}$ and the NS tidal disruption radius $R_{\rm tidal}$ \citep[e.g.,][]{kyutoku2011,shibata2011,foucart2012}. The former depends on the primary BH mass $M_{\rm BH}$ and the dimensionless spin parameter projected onto the orientation of orbital angular momentum $\chi_{\rm BH}$. One can give the normalized ISCO radius $\widetilde{R}_{\rm ISCO} = c^2 R_{\rm ISCO}/GM_{\rm BH}$ \citep{bardeen1972}, i.e., $\widetilde{R}_{\rm {ISCO}} = 3 + Z_2 - {\rm {sign}}(\chi_{\rm {BH}})\sqrt{(3 - Z_1)(3 + Z_1 + 2Z_2)},$ with $Z_1 = 1 + (1 - \chi_{\rm {BH}}^2) ^ {1 / 3} [(1 + \chi_{\rm {BH}})^{1 / 3} + (1 - \chi_{\rm {BH}})^{1 / 3}]$ and $Z_2 = \sqrt{3 \chi_{\rm {BH}}^2 + Z_1^2}$. The radius at which tidal disruption occurs, in the Newtonian approximation, can be expressed as $R_{\rm tidal}\sim R_{\rm NS}(3M_{\rm BH}/M_{\rm NS})^{1/3}$ which is a function of BH mass $M_{\rm B H}$, NS mass $M_{\rm NS}$, and NS radius $R_{\rm NS}$. For a NS with a certain mass $M_{\rm NS}$, its radius $R_{\rm NS}$ and compactness $C_{\rm NS} = GM_{\rm NS}/c^2R_{\rm NS}$ are dependent on the NS equation of state (EoS).

When a NS is tidally disrupted, numerical relativity (NR) simulations \citep[e.g.,][]{foucart2013,kyutoku2013,kyutoku2015} showed that plenty of materials would remain outside the remnant BH to form a disk and an unbound tidal tail. In order to judge whether or not tidal disruption happens for a GW event, evaluation of the amount of total baryon mass after NSBH mergers could be an important quantitative diagnostic. \cite{foucart2018}(hereafter \defcitealias{foucart2018}{F18}\citetalias{foucart2018}) adopted an empirical model to estimate the total amount of remnant masses $M_{\rm total,fit}$ outside the BH horizon as a non-linear function of ISCO radius and tidal disruption radius with the consideration of 75 NR simulations. The model of \citetalias{foucart2018} gives 
\begin{equation}
\label{equ:TotalEjectaMassFunction}
    \frac{M_{\rm total,fit}}{M^{\rm b}_{\rm NS}} = \left[\max\left(\alpha\frac{1 - 2C_{\rm NS}}{\eta^{1 / 3}} - \beta \widetilde{R}_{\rm {ISCO}}\frac{C_{\rm NS}}{\eta} + \gamma , 0 \right)\right]^ {\delta},
\end{equation}
where $\alpha = 0.406$, $\beta = 0.139$, $\gamma = 0.255$, $\delta = 1.761$, $ M^{\rm b}_{\rm NS}$ is the baryonic mass of the NS, $\eta = Q / (1 + Q) ^ 2$, and $Q = M_{\rm BH} / M_{\rm NS}$ is the mass ratio between the BH mass and the NS mass. The more accurate application range of this formula is $Q \in [1 , 7]$, $\chi_{\rm BH} \in [-0.5 , 0.9]$, and $C_{\rm NS} \in [0.13 , 0.182]$ \citep{foucart2018}. A zero $M_{\rm total,fit}$ corresponds to no tidal disruption, which means that the NSBH merger is a plunging event.

Similarly, based on \citetalias{foucart2018}, \cite{zhu2020a} (hereafter \defcitealias{zhu2020a}{Z20}\citetalias{zhu2020a}) presented a similar formula for tidal dynamical ejecta mass by considering 66 NR simulations. The fitting parameters are $\alpha = 0.218$, $\beta = 0.028$, $\gamma = -0.122$, and $\delta = 1.358$. The formula covers the accurate range of $Q \in [1 , 7]$, $\chi_{\rm BH} \in [-0.05 , 0.9]$, and $C_{\rm NS} \in [0.108 , 0.18]$. In addition, \cite{kawaguchi2016} (hereafter \defcitealias{kawaguchi2016}{K16}\citetalias{kawaguchi2016}) and \cite{kruger2020} (hereafter \defcitealias{kruger2020}{KF20}\citetalias{kruger2020}) also provided fitting formulas to model the dynamical ejecta mass. 

We note that the fittings of NR simulations ignored the effects of precession, NS spin, and orbital eccentricity. Since the fitting formulas of the total remnant mass and the dynamical ejecta mass are obtained with independent NR data, one needs to set an upper limit on the maximum fraction of dynamical ejecta mass to the total remnant mass, i.e., $M_{\rm d,max} = f_{\rm max}M_{\rm total,fit}$. We set $f_{\rm max} \approx 0.5$ based on NR simulation results \citep{kyutoku2015}. Therefore, the final empirical mass of the dynamical ejecta is $M_{\rm d} = \min(M_{\rm d,fit} , f_{\rm max}M_{\rm total,fit})$.

\subsection{Source Properties and Tidal Disruption Probability}

\begin{deluxetable*}{ccccc}[htpb!]
\tablecaption{Source properties for potential NSBH events\label{tab:Samples}}
\tablecolumns{5}
\tablewidth{0pt}
\tablehead{
\colhead{GW Event} &
\colhead{GW190426} &
\colhead{GW190814} &
\colhead{GW200105} &
\colhead{GW200115}
}
\startdata
Primary mass $M_1 / M_\odot$ & $5.7^{+3.9}_{-2.3}$ & $23.2^{+1.1}_{-1.0}$ & $8.9^{+1.1}_{-1.3}$ & $5.9^{+1.4}_{-2.1}$  \\
Secondary mass $M_2 / M_\odot$ & $1.5^{+0.8}_{-0.5}$ & $2.59^{+0.08}_{-0.09}$ & $1.9^{+0.2}_{-0.2}$ & $1.4^{+0.6}_{-0.2}$ \\
Mass ratio $Q = M_1 / M_2 $ & $4.2^{+6.7}_{-2.7}$ & $8.9^{+0.8}_{-0.6}$ & $4.8^{+1.1}_{-1.1}$ & $4.2^{+2.1}_{-2.3}$ \\
Effective inspiral spin $\chi_{\rm eff}$ & $-0.03^{+0.32}_{-0.30}$ & $-0.002^{+0.060}_{-0.061}$ & $-0.01^{+0.08}_{-0.12}$ & $-0.14^{+0.17}_{-0.34}$ \\
Luminosity distance $D_{\rm L} / {\rm Mpc}$ & $370^{+190}_{-160}$ & $241^{+41}_{-45}$ & $280^{+110}_{-110}$ & $310^{+150}_{-110}$
\enddata
\tablecomments{Combined source properties of four potential NSBH events inferred from low-spin priors of the secondary component. We report the median values with $90\%$ credible intervals. \\References: GW190426 \citep{abbott2020gwtc2}, GW190814 \citep{abbott2020GW190814}, GW200105, and GW 200115 \citep{abbott2021NSBH}.}
\end{deluxetable*}

\begin{figure*}[htpb]
    \centering
    \includegraphics[width = 0.49\linewidth , trim = 70 30 70 30, clip]{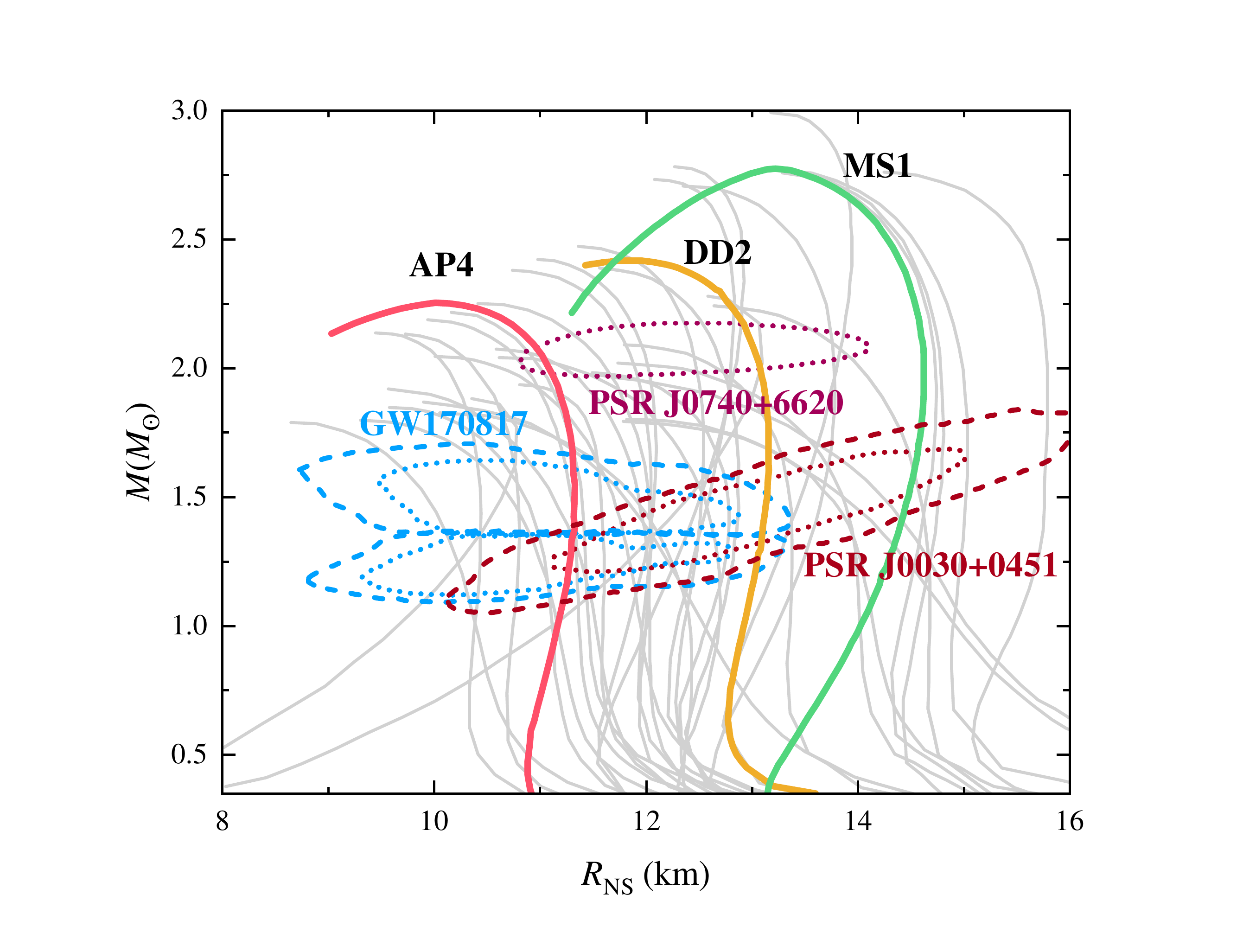}
    \caption{{Observational constraints on the mass-radius relationship of selected sources, compared with NS EoS models. The blue, dark violet, and carmine are observational constraints by the GW170817 \citep[][dotted contours at 68\% and dashed contours at 95\% confidence levels]{abbott2018measurement}, NICER measurement of PSR\,J0740+6620 \citep[][68\% confidence level]{miller2021}, and NICER measurement of PSR\,J0030+0451 \citep[][dotted contours at 68\% and dashed contours at 95\% confidence levels]{riley2019} \citep[see][who used more observational limits to constrain the mass-radius relationship]{dietrich2020}. Colored solid lines are the three selected representative EoSs used in this work. Other EoSs (gray solid lines) are obtained from \cite{bauswein2012,bauswein2013}. }}
    \label{fig:EoSs}
\end{figure*}

The physical properties of the four GW events using the methodology of a coherent Bayesian analysis \citep{abbott2019gwtc1} have been released recently. It is plausibly expected that most NSs would have low spins before NSBH mergers since NSs would have spun down via magnetic dipole radiation during the long history in the pre-merger phase. This has been suggested by both pulsar observations \citep[e.g.,][]{manchester2005} and population synthesis models \citep[e.g.,][]{kiel2008,osowski2011}. Therefore, we only collect the posterior samples of binary parameters with consideration of a low-spin prior assumption. The posterior results of these four events are shown in Table \ref{tab:Samples}. In order to calculate the tidal disruption probability, we need to know $\chi_{\rm BH}$ and the EoS. We use the primary's aligned spin components (i.e., $\chi_{1z}$) in the posterior samples to estimate $\chi_{\rm BH}$. For the NS EoS, these four potential NSBH GW events have poor measurements for tidal deformability \citep{abbott2021NSBH} so that EoS cannot be constrained directly.
{\cite{abbott2021NSBH} adopted the method presented by \cite{stachie2021} to marginalize over the EoSs when calculating the ejecta mass for GW200105 and GW200115, and showed that their ejecta masses are essentially negligible.} In our work, we select three specific {representative} EoSs from soft to stiff {as shown in Figure \ref{fig:EoSs}}, including AP4 \citep{akmal1997}, DD2 \citep{typel2010}, and Ms1 \citep{muller1996}. Among them, AP4 is one of the most likely EoS while DD2 is permitted as one of the stiffest EoS constrained by GW170817 \citep{abbott2018measurement,abbott2019properties} {and \cite{dietrich2020}}. One can thus calculate the NS barynoic mass $M_{\rm NS}^{\rm b}$ as a function of $M_{\rm NS}$ introduced by \cite{gao2020}:
\begin{equation}
\label{Eq: CalculateBaryonicMass}
    M_{\rm NS}^{\rm b} = M_{\rm NS} + A_1\times M_{\rm NS}^2 + A_2\times M_{\rm NS}^3,
\end{equation}
where $A_1$ and $A_2$ for each selected EoS are presented in Table \ref{tab:EoSs}, and $M_{\rm NS}^{\rm b}$ and $M_{\rm NS}$ in this formula are in units of $M_\odot$. We estimate the compactness of NS based on the empirical formula given by \cite{coughlin2017}
\begin{equation}
    C_{\rm NS} =  1.1056\times(M_{\rm NS}^{\rm b} / M_{\rm NS} - 1)^{0.8277}.
\end{equation}

\begin{deluxetable}{cccc}[tpb!]
\tablecaption{Characteristic Parameters for the Selected EoSs\label{tab:EoSs}}
\tablecolumns{4}
\tablewidth{0pt}
\tablehead{
\colhead{EoS} &
\colhead{$M_{\rm TOV}/M_\odot$} &
\colhead{$A_1$} &
\colhead{$A_2$}
}
\startdata
AP4 & 2.22 & 0.045 & 0.023 \\
DD2 & 2.42 & 0.046 & 0.014 \\
Ms1 & 2.77 & 0.042 & 0.010
\enddata
\tablecomments{The columns from left to right represent the NS EoSs we selected, the NS TOV mass $M_{\rm TOV}$, and best fit values of $A_1$ and $A_2$ for each EoS in Equation (\ref{Eq: CalculateBaryonicMass}). All the values of each parameter are cited from \cite{gao2020}.}
\end{deluxetable}

\begin{figure*}[htpb]
    \centering
    \includegraphics[width = 0.49\linewidth , trim = 50 10 70 30, clip]{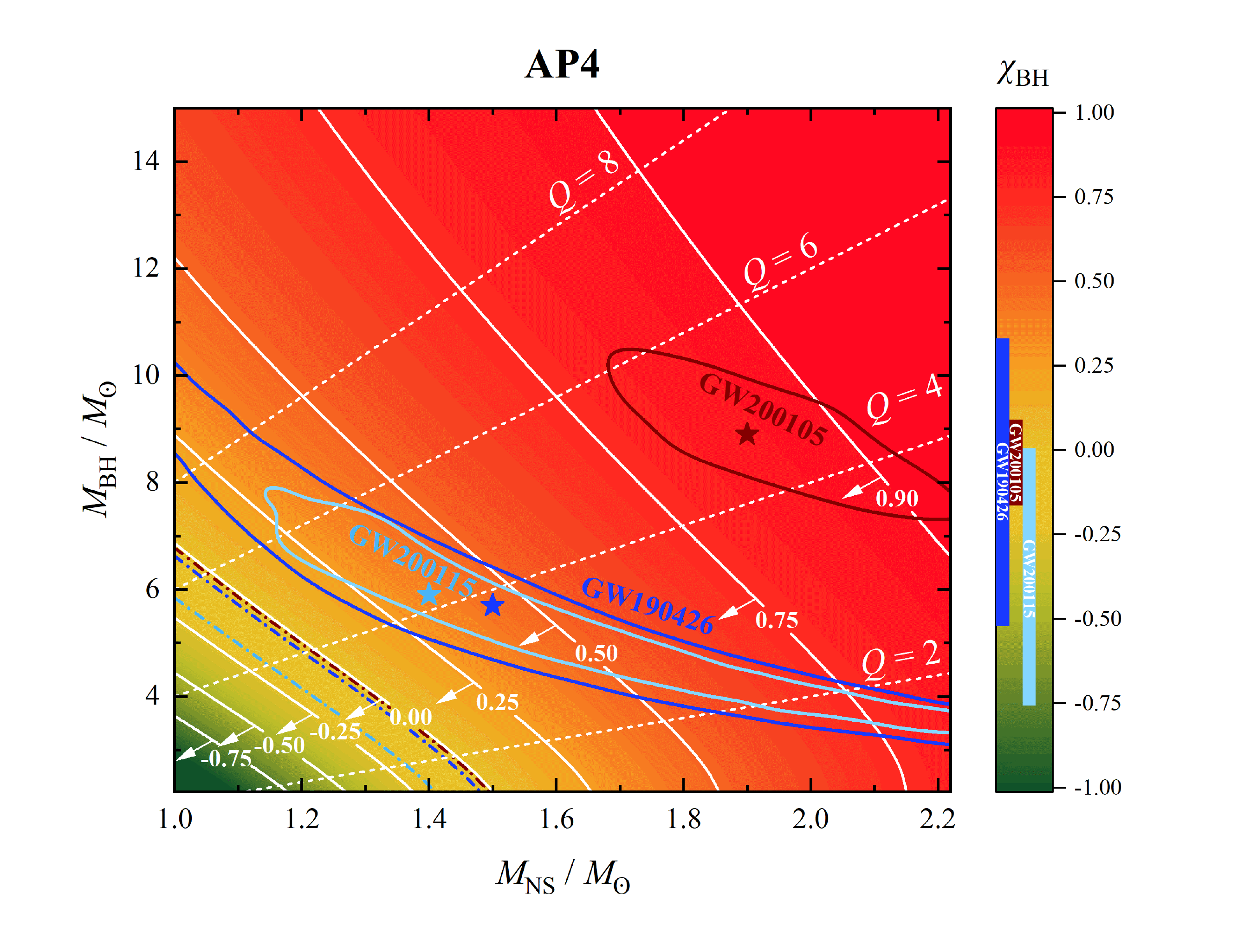}
    \includegraphics[width = 0.49\linewidth , trim = 50 10 70 30, clip]{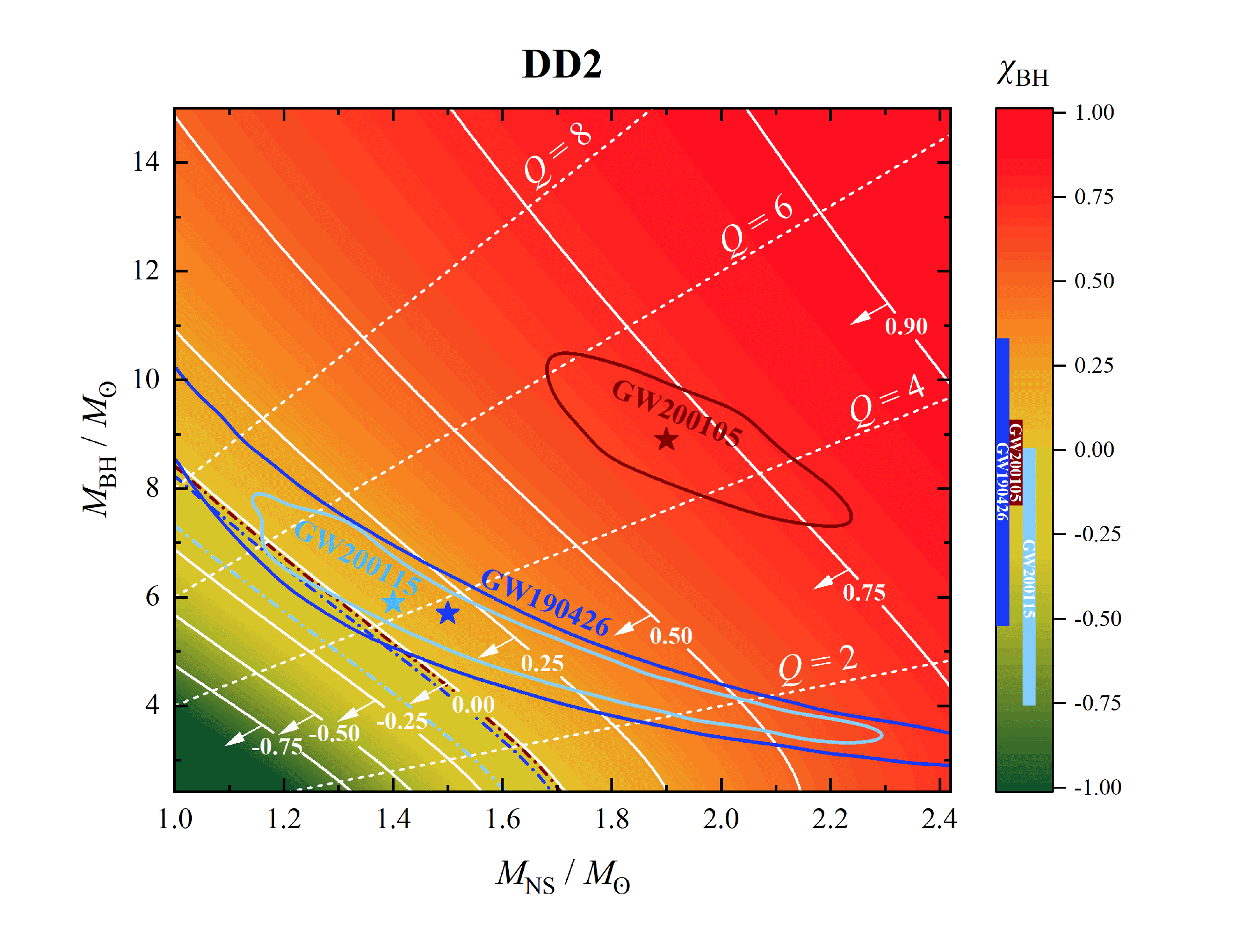}
    \includegraphics[width = 0.49\linewidth , trim = 50 30 70 30, clip]{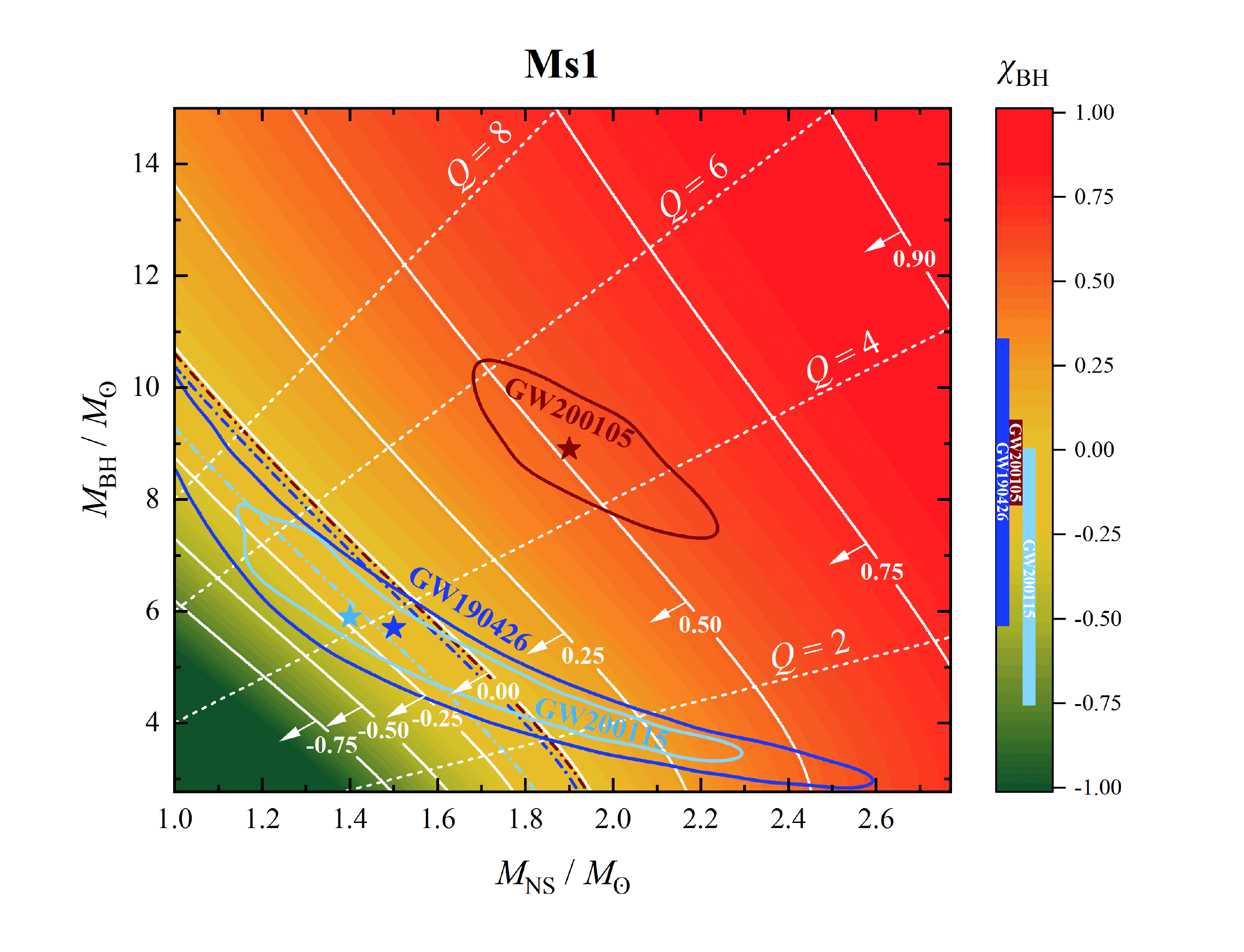}
    \caption{The source-frame mass parameter space for NSBH merger systems to allow tidal disruption of the NS by the BH. Three EoSs are considered: AP4 (top left panel), DD2 (top right panel), and Ms1 (bottom panel). The dashed lines represent mass ratio $Q = M_{\rm BH} / M_{\rm NS}$ from $Q = 2$ to $Q = 8$. We mark several values of primary BH spin along the orbital angular momentum from $\chi_{\rm BH} = - 0.75$ to $\chi_{\rm BH} = 0.90$ as solid lines in each panel. For a specific $\chi_{\rm BH}$, the NSBH mergers with component masses located at the bottom left parameter space (denoted by the direction of the arrows) can allow tidal disruptions to occur. For GW190426 (dark blue), GW200105 (brown), and GW200115 (light blue), the $90\%$ credible posterior distributions (colored solid lines) and the medians (colored star points) are displayed. Corresponding median values of $\chi_{\rm BH}$ for these three sources are marked as dashed-dotted lines while their $90\%$ posterior distributions are labeled at the dextral color bars of each panel.}
    \label{fig:TidalDisruptionRegion}
\end{figure*}

Figure \ref{fig:TidalDisruptionRegion} shows the parameter space where the NS can be tidally disrupted using the \citetalias{foucart2018} model. The tidal disruption tends to occur if the NSBH binaries have a low-mass NS component with a stiff EoS, and a high-spin low-mass BH component. In particular, the parameter space would significantly increase if the primary BH has a larger $\chi_{\rm BH}$ value. If the BH component carries a low spin or even has an opposite spin with respect to the  orbital angular momentum (like O3 LVC NSBH candidates), the mass space that allows tidal disruption of the NS would be limited.

We also show the posterior distribution for the component masses of the O3 LVC NSBH candidates in Figure \ref{fig:TidalDisruptionRegion}. The explicit results of tidal disruption probabilities for these four events are collected in Table \ref{tab:MassDistribution}. It is obvious that there is no probability for GW190814 (if it is a NSBH) and GW200105 to have tidal disruption, so no EM signals are expected following these two merger events. The probability for GW200115 to have material outside of the remnant BH horizon is also small, unless the EoS of NS component is extremely stiff. GW190426 is the most likely event for tidal disruption. In view that DD2 is one of the stiffest EoS allowed by the observations of GW170817 \citep{abbott2019properties}, the probability could be smaller than $\lesssim 24\%$.

\begin{deluxetable*}{cccc|ccc}[tb!]
\tablecaption{Tidal disruption probability and dynamical ejecta mass distribution\label{tab:MassDistribution}}
\tablecolumns{7}
\tablewidth{0pt}
\tablehead{
\colhead{GW Event} &
\colhead{EoS} &
\colhead{$P_{\rm NSBH}$\tablenotemark{a}} &
\colhead{Tidal Disruption Probability} &
\multicolumn{3}{c}{Dynamical Ejecta Mass\tablenotemark{b}} 
}
\startdata
{} & {} & {} & \citetalias{foucart2018} & \citetalias{kawaguchi2016} & \citetalias{kruger2020} & \citetalias{zhu2020a} \\
\hline
\multirow{3}{*}{GW190426} & AP4 & $94.4\%$  & $5.95\%$ & $1.9^{+6.1}_{-1.8}\times10^{-3}M_\odot$ & $5.3^{+8.7}_{-4.8}\times10^{-3}M_\odot$ & $1.7^{+6.2}_{-1.7}\times10^{-3}M_\odot$ \\
{} & DD2 & $97.6\%$ & $24.3\%$ & $7^{+16}_{-6}\times10^{-3}M_\odot$ & 
$10^{+14}_{-9}\times10^{-3}M_\odot$ &
$5^{+17}_{-5}\times10^{-3}M_\odot$ \\
{} & Ms1 & $99.8\%$ & $65.2\%$ & $1.5^{+3.3}_{-1.3}\times10^{-2}M_\odot$ & $1.5^{+3.4}_{-1.2}\times10^{-2}M_\odot$ & $1.3^{+3.4}_{-1.2}\times10^{-2}M_\odot$ \\
\hline
\multirow{3}{*}{GW190814} & AP4 & $0\%$ & -- &  -- & -- & -- \\
{} & DD2 & $0.30\%$ & $0\%$ & $0$ & $0$ & $0$  \\
{} & Ms1 & $99.9\%$ & $0\%$ & $0$ & $0$ & $0$  \\
\hline
\multirow{3}{*}{GW200105} & AP4 & $97.0\%$  & $0\%$ & $0$ & $0$ & $0$ \\
{} & DD2 & $99.1\%$ & $0\%$ & $0$ & $0$ & $0$ \\
{} & Ms1 & $99.8\%$ & $0\%$ & $0$ & $0$ & $0$ \\
\hline
\multirow{3}{*}{GW200115} & AP4 & $98.1\%$ & $0\%$ & $0$ & $0$ & $0$ \\
{} & DD2 & $100\%$ & $2.76\%$ & $6^{+39}_{-6}\times10^{-4}\,M_\odot$ &$34^{+41}_{-33}\times10^{-4}\,M_\odot$ &
$6^{+39}_{-6}\times10^{-4}\,M_\odot$ \\
{} & Ms1 & $100\%$ & $49.9\%$ & $6^{+11}_{-6}\times10^{-3}\,M_\odot$ &
$7^{+11}_{-5}\times10^{-3}\,M_\odot$ &
$6^{+11}_{-6}\times10^{-3}\,M_\odot$ 
\enddata
\tablenotetext{a}{The probability that the GW event is a NSBH merger, i.e., the primary mass $M_1 > M_{\rm TOV}$ and the secondary mass $M_2\leq M_{\rm TOV}$.}
\tablenotetext{b}{Median values with 90\% credible intervals of dynamical ejecta mass are calculated for events parameter space that can be tidally disrupted.}
\tablecomments{References: (1) \citetalias{foucart2018} \citep{foucart2018}; (2) \citetalias{kawaguchi2016} \citep{kawaguchi2016}; (3) \citetalias{kruger2020} \citep{kruger2020}; (4) \citetalias{zhu2020a} \citep{zhu2020a}.}
\end{deluxetable*}

\begin{figure*}[htpb]
    \centering
    \includegraphics[width = 0.49\linewidth , trim = 70 28 90 30, clip]{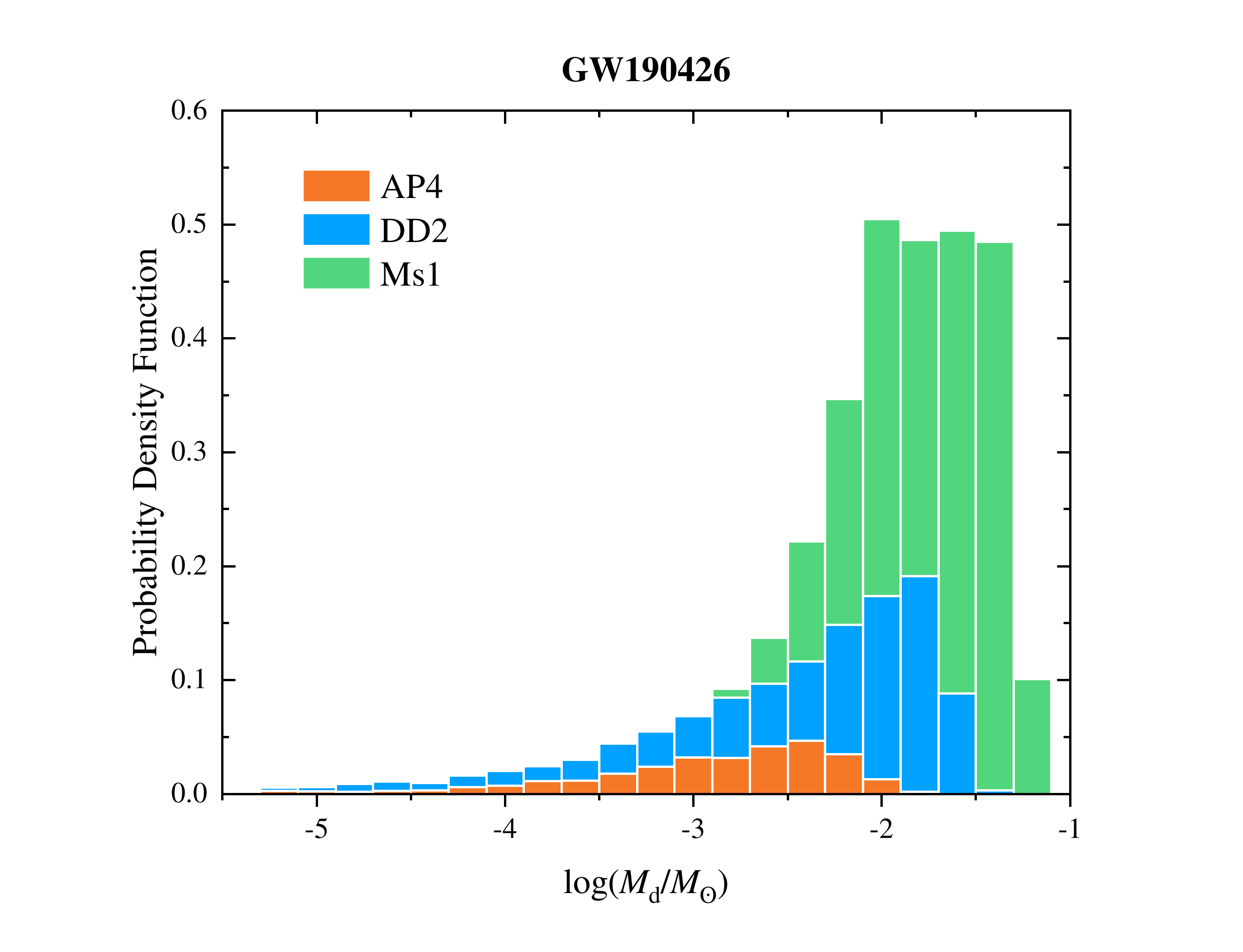}
    \includegraphics[width = 0.49\linewidth , trim = 70 28 90 30, clip]{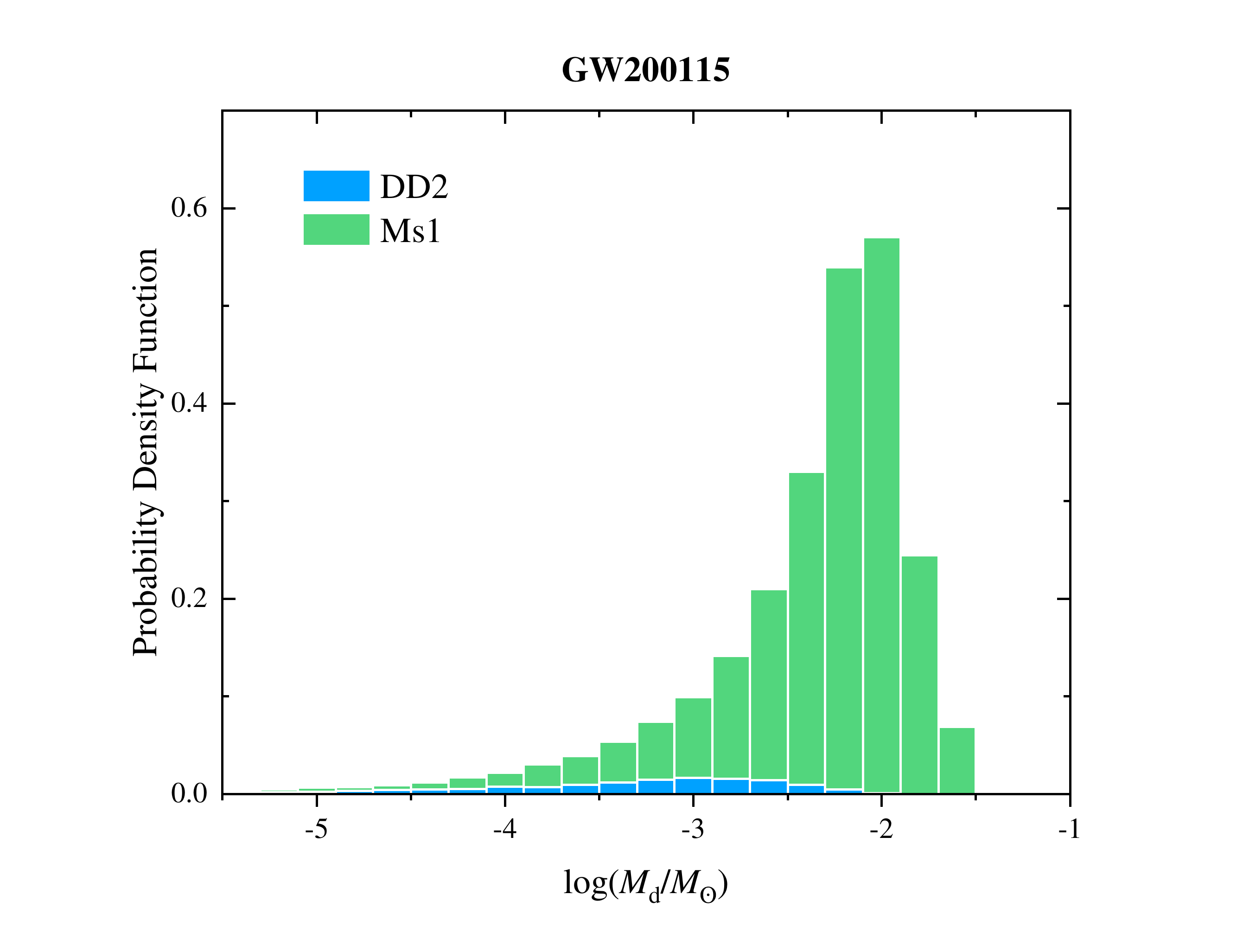}
    \caption{Probability density distributions of the dynamical ejecta mass for GW190426 (left panel) and GW200115 (right panel). The orange, blue, and green histograms represent the probability density of dynamical ejecta mass with the consideration of the EoSs AP4, DD2, and Ms1, respectively. The bin width of the histograms is set as $\Delta = 0.2$ in logarithmic scale.}
    \label{fig:EjectaMassDisribution}
\end{figure*}

\subsection{Dynamical Ejecta Mass and Luminosity Distributions}

\cite{zhu2020a} showed that the fastest moving lanthanide-rich dynamical ejecta usually contribute to the majority of the NSBH kilonova emission, because the BH-torus systems formed after NSBH mergers are usually hard to produce a large amount of lanthanide-free ejecta \citep[e.g.,][]{fernandez2013,just2015,siegel2017}. We  calculate the amount of dynamical ejecta mass for the tidal disruption parameter space of these four NSBH GW events (see Table \ref{tab:MassDistribution}) and give the probability density distributions of the dynamical ejecta for GW190426 and GW200115 using \citetalias{zhu2020a} model as shown in Figure \ref{fig:EjectaMassDisribution}. Comparing with \citetalias{kawaguchi2016} and \citetalias{zhu2020a} models, the \citetalias{kruger2020} model predictes a relatively larger value for the dynamical ejecta but the difference is not too significant. The mass of the dynamical ejecta estimated by  fitting the GW170817/AT\,2017gfo lightcurve mainly lies in the range of $(0.01-0.05)\,M_\odot$ \citep[e.g.,][]{cowperthwaite2017,kasen2017,kasliwal2017,perego2017,tanaka2017,villar2017}. Only GW190426 has a chance to generate a similar amount of dynamical ejecta with GW 170817/AT\,2017gfo if it has a NS companion with the Ms1 EoS. If the NS has an EoS of AP4 or DD2 which were not ruled out by GW170817 \citep{abbott2019properties}, the generated dynamical ejecta masses of these two NSBH GW events should be $\lesssim\sim 10^{-2}\,M_\odot$. Even if GW190426 and GW200115 can make tidal disruption, these results indicate that the associated kilonovae would be dim.

\begin{figure*}[htpb]
    \centering
    \includegraphics[width = 0.49\linewidth , trim = 70 10 90 30, clip]{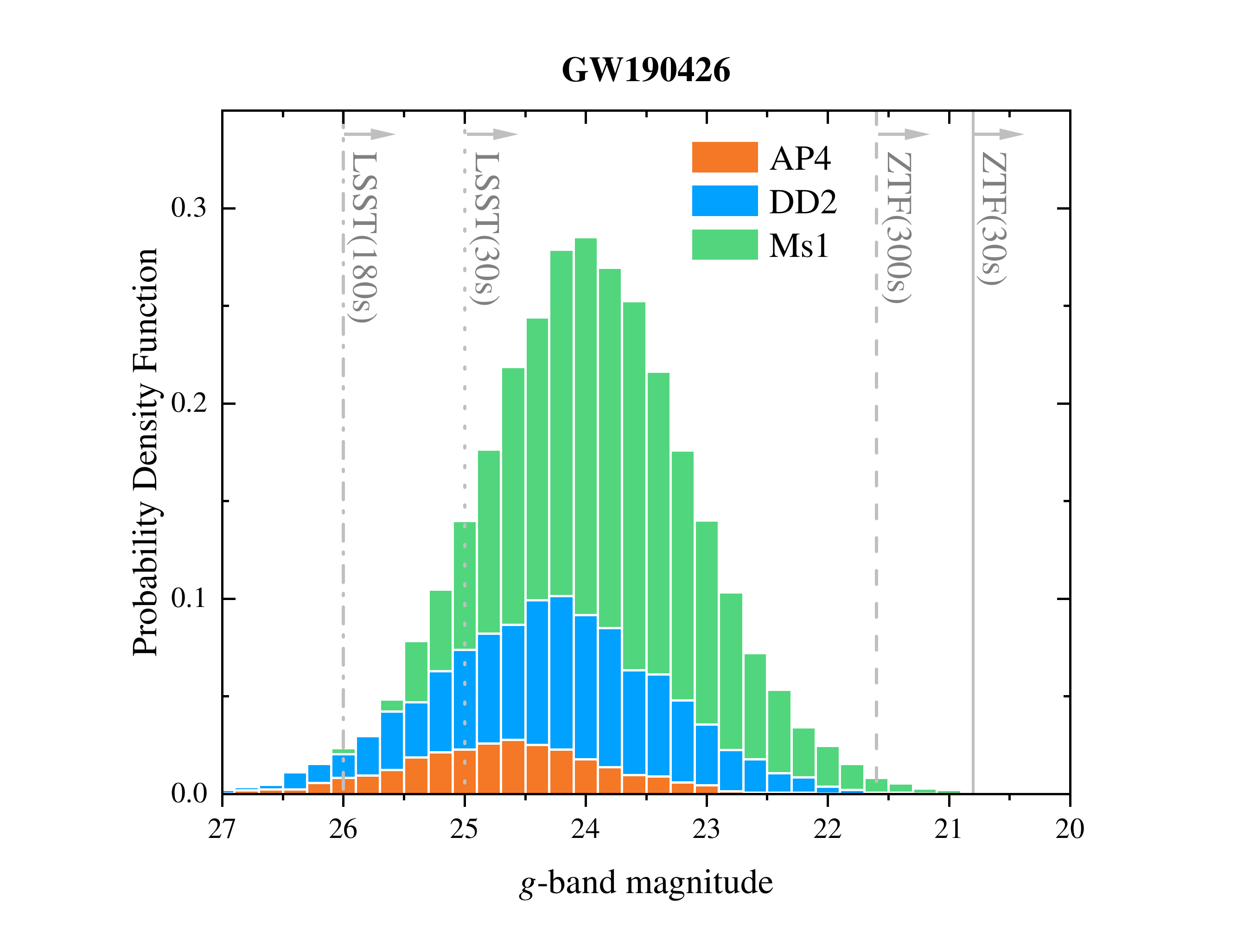}
    \includegraphics[width = 0.49\linewidth , trim = 70 10 90 30, clip]{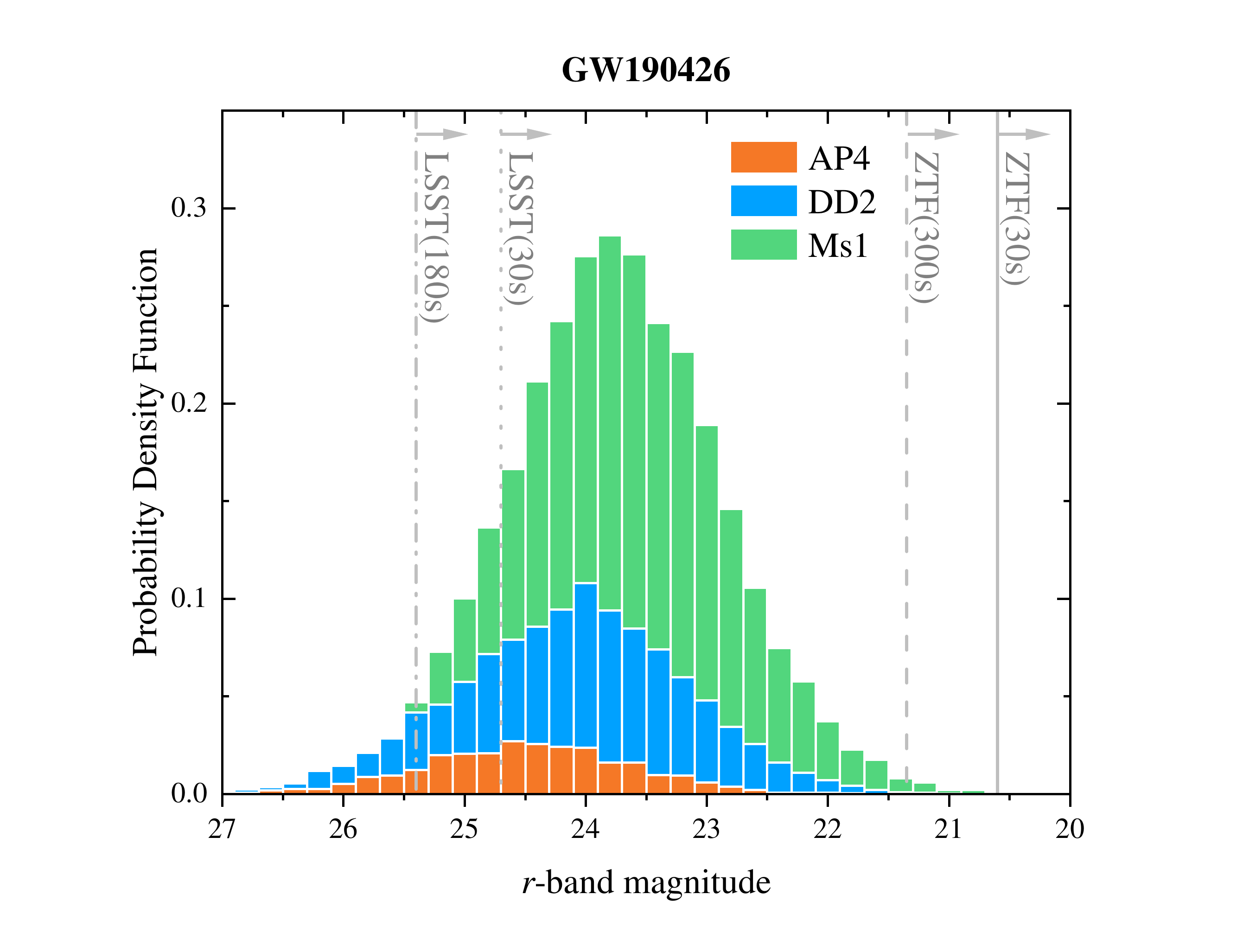}
    \includegraphics[width = 0.49\linewidth , trim = 70 30 90 30, clip]{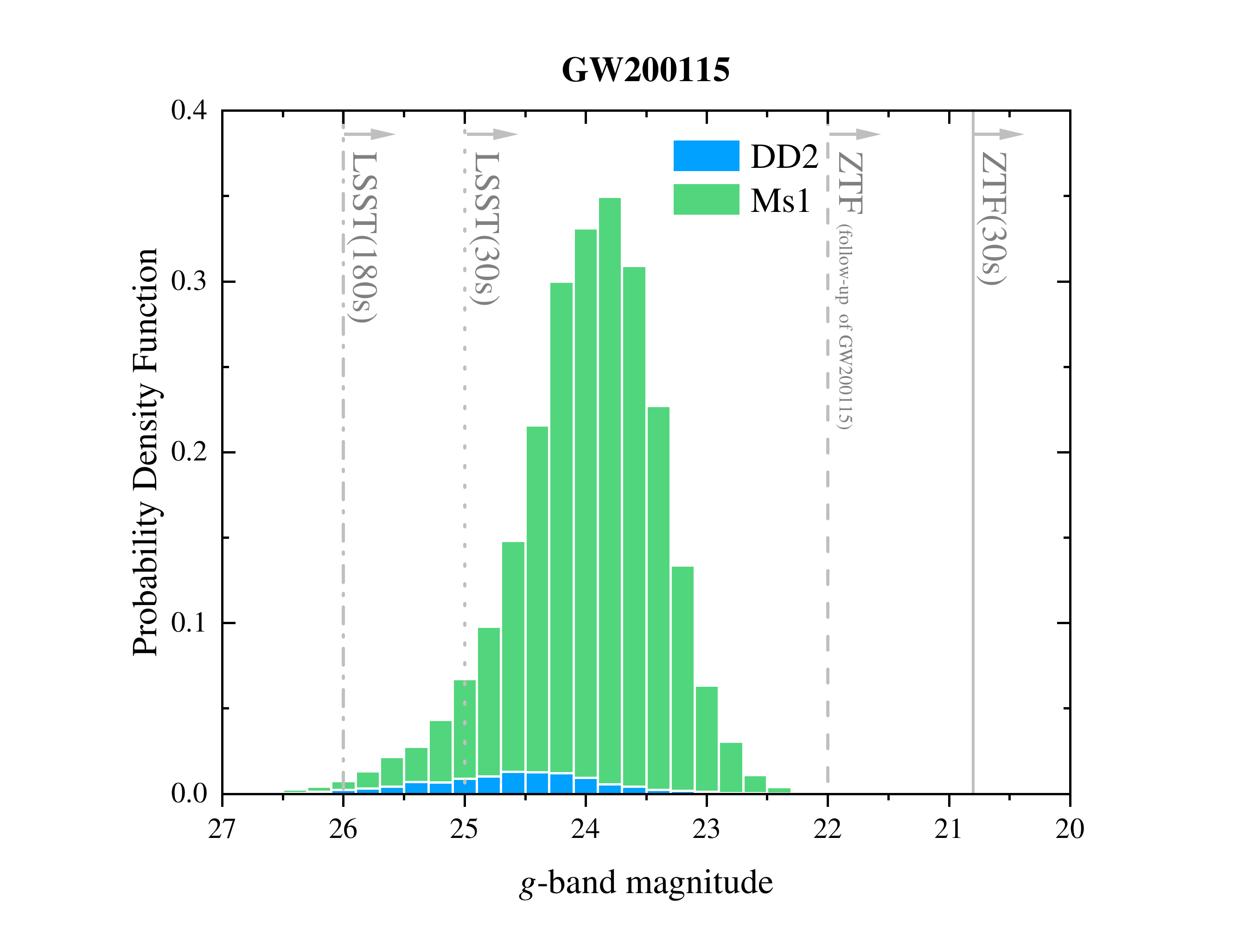}
    \includegraphics[width = 0.49\linewidth , trim = 70 30 90 30, clip]{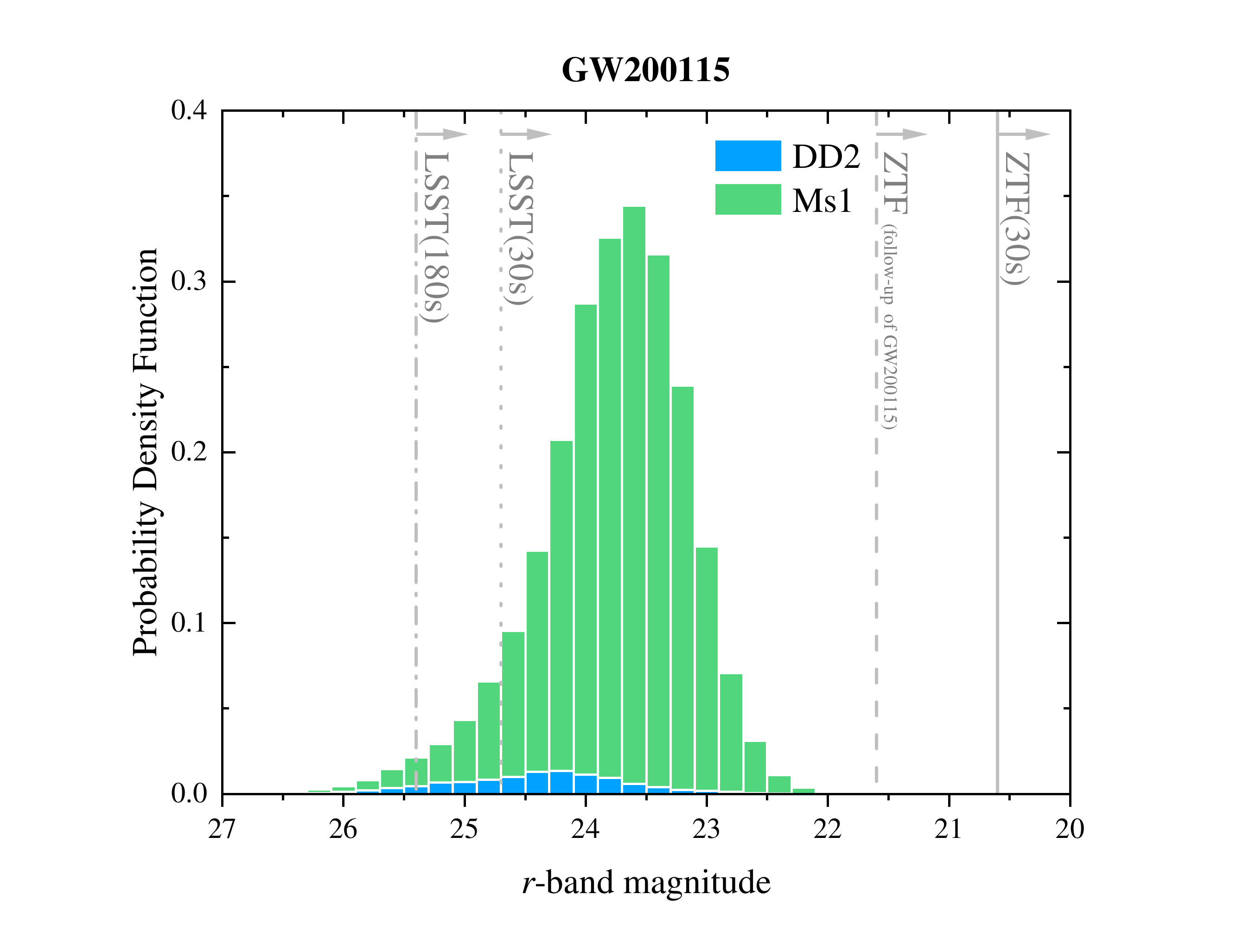}
    \caption{Probability density distributions of NSBH merger kilonova $g$-band peak magnitude (left panels) and $r$-band peak magnitude (right panels) for GW190426 (top panels) and GW200115 (bottom panels). The orange, blue, and green histograms represent the probability density of kilonova peak magnitude with the consideration of the EoSs AP4, DD2, and Ms1, respectively. The gray solid, dotted, and dashed-dotted lines show the $5\sigma$ ZTF normal survey depth, the LSST normal survey depth, and the LSST threshold depth for exposure time of $180\,{\rm s}$ under the ideal observing conditions. The ZTF deepest magnitudes for follow-up searches of {GW190426 and} GW200115, respectively reported by \cite{coughlin2019,anand2020}, are marked by dashed lines. Here, the bin width of the histograms is set as $\Delta = 0.2$. }
    \label{fig:LuminosityDistribution}
\end{figure*}

By taking advantage of the viewing-angle dependent NSBH merger kilonova model from \cite{zhu2020a}, we show the probability density distributions of $g$-band and $r$-band peak magnitudes for GW190426 and GW200115 in Figure \ref{fig:LuminosityDistribution}. At present, the Zwicky Transient Facility \citep[ZTF;][]{graham2019} possess one of the most powerful capabilities for GW-triggered follow-up observations. The deepest $5\sigma$ observed depths of ZTF for GW200105 and GW200115 follow-up observations are $m_{\rm AB} \approx 22\,{\rm mag}$ reported by \cite{anand2020}. However, even though these two events could produce kilonovae, it is impossible for ZTF to observe them regardless of which EoS is adopted. We also mark the detection depths of Large Synoptic Survey Telescope \citep[LSST;][]{lsst2009} with $30\,{\rm s}$ and $180\,{\rm s}$ exposures in Figure \ref{fig:LuminosityDistribution}. LSST can have an enough detectability to discover the kilonova emission from GW190426 and GW200115 if the NS EoS is stiff so that the component NSs can be tidally disrupted.

\section{Implications from Population Synthesis Results}

\begin{figure*}[htpb]
    \centering
    \includegraphics[width = 0.49\linewidth , trim = 50 10 70 20, clip]{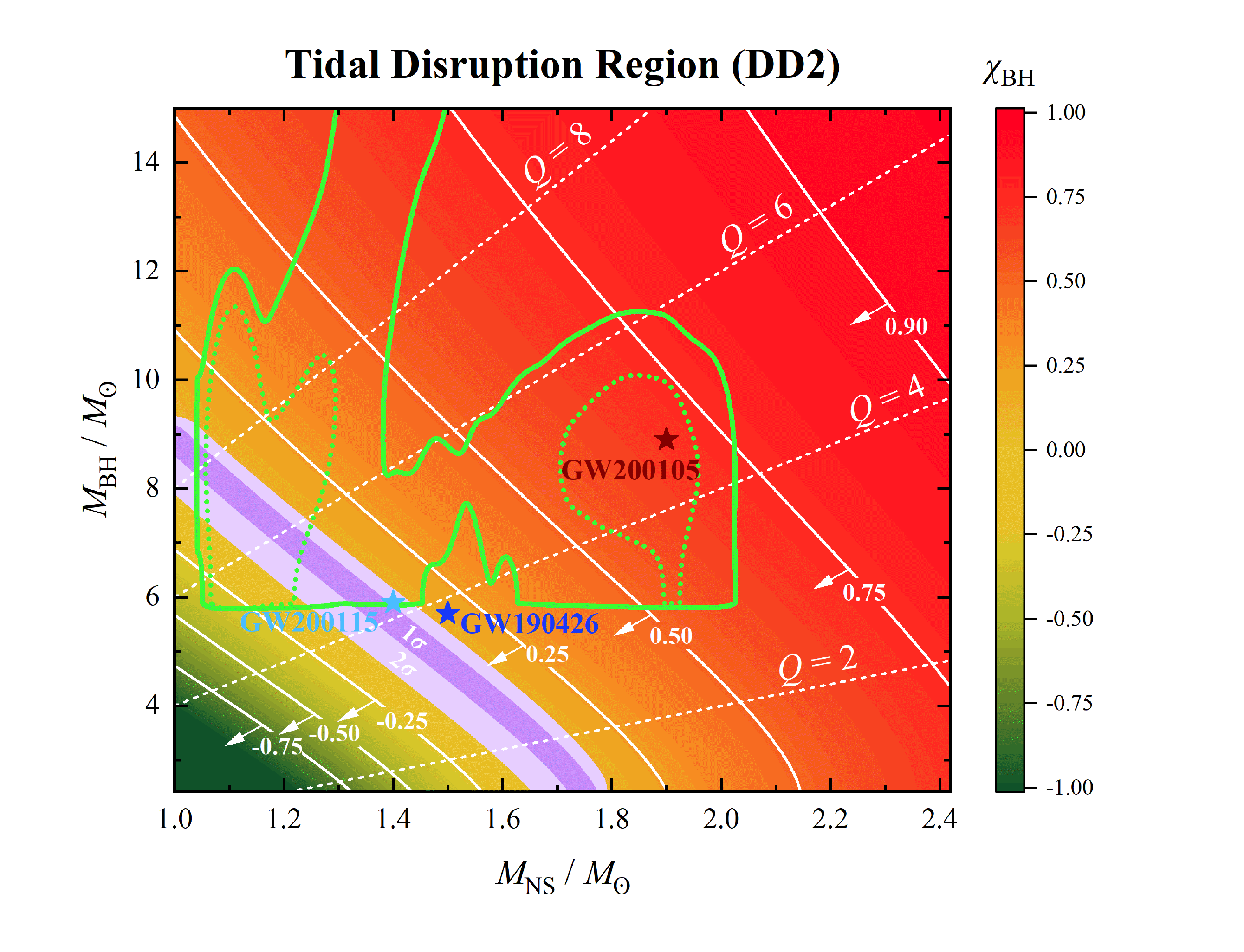}
    \caption{Similar to Figure \ref{fig:TidalDisruptionRegion} but for source-frame mass parameter space where tidal disruption can occur by considering the specific NS EoS DD2. The green dotted and solid lines represent $1\sigma$ and $2\sigma$ source-frame masses distributions of population synthesis simulations from \cite{belcynski2020}. The purple and light purple regions represent $1\sigma$ and $2\sigma$ distributions of $\chi_{\rm BH}$.}
    \label{fig:population}
\end{figure*}

\begin{figure*}[htpb]
    \centering
    \includegraphics[width = 0.49\linewidth , trim = 50 10 70 20, clip]{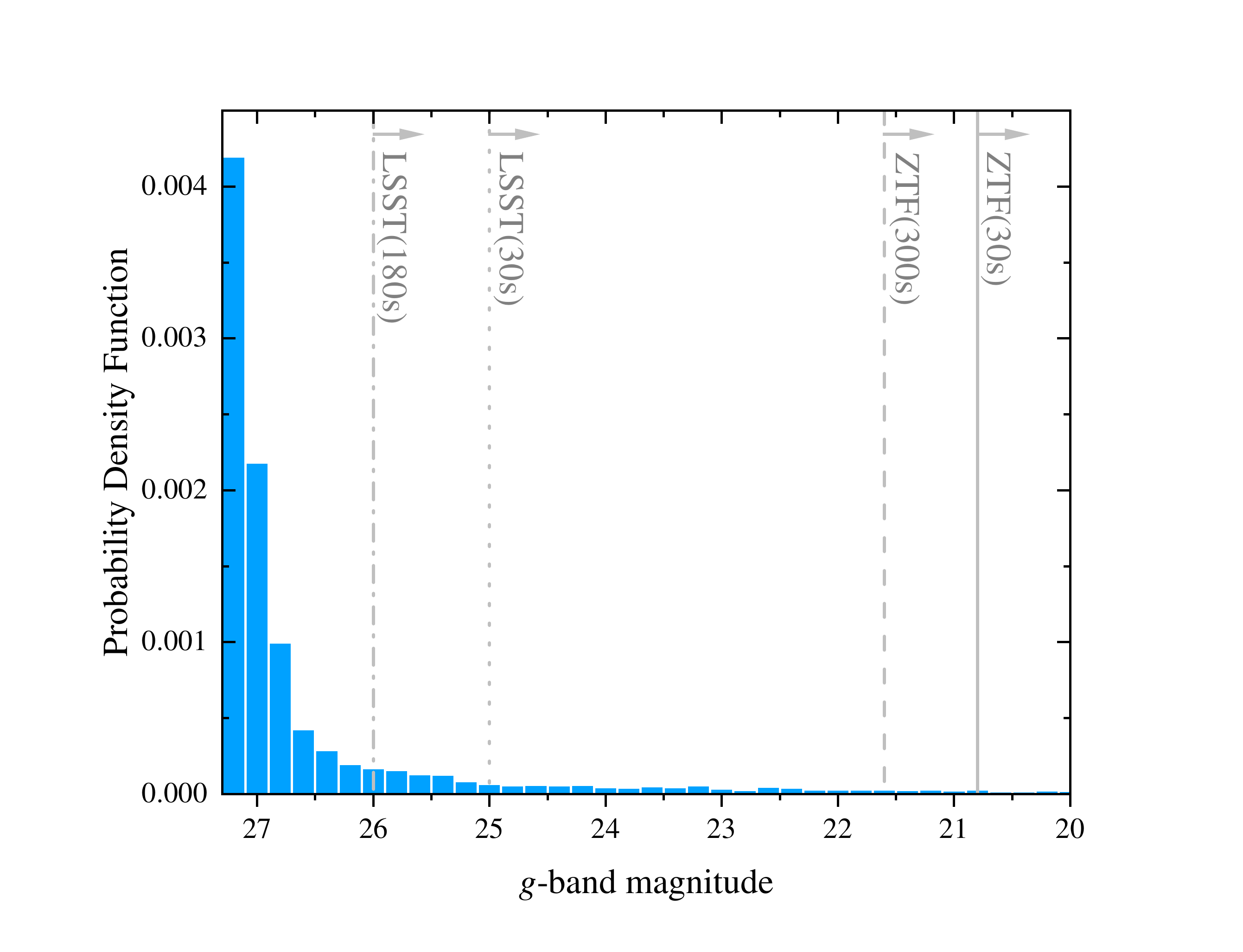}
    \includegraphics[width = 0.49\linewidth , trim = 50 10 70 20, clip]{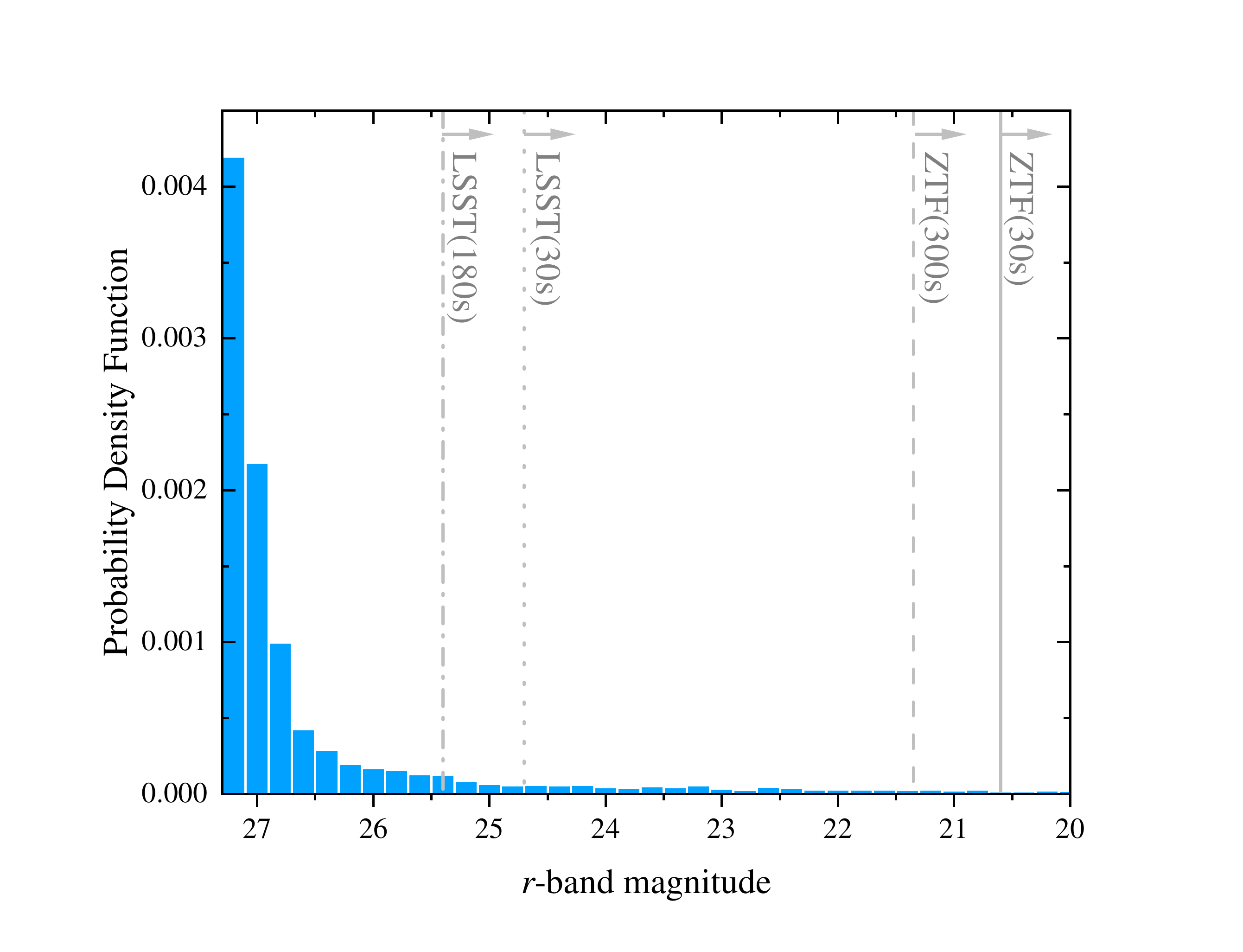}
    \caption{Probability density distributions of $g$-band and $r$-band kilonova peak apparent magnitudes for the simulated cosmological NSBH mergers. The gray solid, {dashed,} dotted, and dashed-dotted lines show the $5\sigma$ ZTF normal survey depth, {the ZTF threshold depth for exposure time of $300\,{\rm s}$}, the LSST normal survey depth, and the LSST threshold depth for exposure time of $180\,{\rm s}$ under ideal observing conditions. The bin width of the histograms is set as $\Delta = 0.2$.}
    \label{fig:luminosity function}
\end{figure*}

By considering several models with different common envelope treatments, stellar remnant natal kicks, cosmic chemical evolutions, and angular momentum transport mechanisms during BH formation, \cite{belcynski2020} calculated the event rates for BNS, NSBH, and binary BH (BBH) mergers, and predicted the effective spin parameters of BBH mergers. The most recent local BNS, NSBH, and BBH merger rate densities inferred from GW observations \citep{abbott2020population,abbott2021NSBH} are $R_{\rm BNS} = 320^{+490}_{-240}\,{\rm Gpc}^{-3}\,{\rm yr}^{-1}$, $R_{\rm NSBH} = 45^{+75}_{-33}\,{\rm Gpc}^{-3}\,{\rm yr}^{-1}$ (or $R_{\rm NSBH} = 130^{+112}_{-69}\,{\rm Gpc}^{-3}\,{\rm yr}^{-1}$)\footnote{The former NSBH merger rate density was obtained under the assumption that GW200105 and GW200115 are representatives of the NSBH population, while the latter one was calculated by assuming a broader distribution of the component mass in O3 triggers \citep{abbott2021NSBH}.}, and $R_{\rm BBH} = 24^{+14}_{-9}\,{\rm Gpc}^{-3}\,{\rm yr}^{-1}$, respectively. The most favored population synthesis models which can match the observed merger event rates are ${\rm M33.A}$ and ${\rm M43.A}$ models \citep[see Figure 24 and Figure 25 in ][]{belcynski2020}. The difference of these two models is due to the different assumptions of angular momentum transport mechanism during BH formation: an efficient transport by the Tayler-Spruit magnetic dynamo \citep[MESA code; see also][]{qin2018} for the ${\rm M33.A}$ model, and a very-efficient transport \citep{fuller2019} for the ${\rm M43.A}$ model. Both angular momentum transport models are favored to explain near-zero effective spins for the observed BBH mergers \citep{belcynski2020}.  

We combine the population synthesis simulation results of ${\rm M33.A}$ and ${\rm M43.A}$ models\footnote{\url{https://www.syntheticuniverse.org/stvsgwo.php}} \citep{belcynski2020} and show the distributions of the source-frame masses and BH spins in Figure \ref{fig:population}. The results reveal that the most common NSBH mergers have BH mass of $\sim(6-12)\,M_\odot$ and $\chi_{\rm BH}$ of $\sim(-0.2- 0.2)$, while the NSs in these systems have a wide range of mass distribution mainly distributed in $\sim(1.1-2.0)\,M_\odot$. These BH and NS mass distributions are similar to other population synthesis results \cite[e.g.,][]{giacobbo2018,broekgaarden2021}. The median values of source-frame masses and BH spins for GW190426, GW200105, and GW200115 are basically located in the $2\sigma$ regions of population synthesis results. By adopting NS EoS of DD2, we find that only $\sim20\%$ cosmological NSBH mergers which have less-massive BHs and less-massive NSs can allow tidal disruption to occur to produce bright kilonovae. We then simulate $5\times10^6$ NSBH mergers in the universe based on the population synthesis simulation results to map the distributions of $r$- and $g$-band observed peak apparent magnitudes (see Figure \ref{fig:luminosity function}). One can conclude that the associated kilonovae for these disrupted events are still  difficult to be discovered by LSST after GW triggers in the future. This is likely because most of NSBH-merger kilonovae have faint brightness and larger distances compared with BNS mergers. Via the Blandford-Znajek mechanism \citep{blandfor1977}, the remnant BH from these mergers can accrete from the surrounding remnant disk for disrupted events and launch a pair of collimated relativistic jets. The sGRB and afterglow would be the more promising EM counterparts of NSBH mergers for detection thanks to the relativistic beaming effect of the jets. Therefore, for future GW-triggered multi-messenger observations, one may search for potential sGRBs and afterglows as plausible EM counterparts for NSBH merger GW events.

\section{Conclusions and Discussion}

In this paper, we have detailedly analyzed the observations of the O3 LVC  NSBH merger candidates and discuss the detectability of the kilonova signals from these events.  The posterior distributions revealed that the NS components in GW190814 (if it is a NSBH merger event) and GW200105 would directly plunge into their respective BH without making any EM counterpart. GW190426 and GW200115 have low probabilities for tidal disruption {which are consistent with the results of LVC \citep{abbott2021NSBH}}, but even if tidal disruptions occurred, the brightness of the associated kilonovae of these two events could be too faint for the present follow-up survey telescopes to detect. This can thus explain why kilonovae were not found by the follow-up observations of these GW events. Considering the NS EoS of DD2 model and the best constrained population synthesis simulation results, we have found that only a small fraction ($\sim20\%$) of cosmological NSBH mergers theoretically can occur tidal disruption and produce kilonova signals. However, the predicted brightness for most of kilonovae still be too faint for the follow-up observations of LSST. This result is basically consistent with other recent studies \cite[e.g.,][]{zappa2019,zhu2020b,drozda2020}. For future GW-triggered multi-messenger observations, one may search for potential sGRB and afterglow as ideal EM counterparts of NSBH GW events.

The kilonova model we used in this work assumes that the BH-torus cannot produce much lanthanide-poor wind ejecta \citep[e.g.,][]{fernandez2013,just2015,siegel2017}. If the electron fraction of the wind ejecta is not really low \citep[e.g.,][]{fujibayashi2020a,fujibayashi2020b}, the NSBH kilonovae can be more luminous. We note that the differences in peak luminosity of NSBH kilonovae between the model of \cite{zhu2020a} and other models \citep[e.g.,][]{kawaguchi2016,kawaguchi2020,barbieri2019,darbha2021} are within a factor of two, corresponding to an uncertainty of $\sim1\,{\rm mag}$ for the peak luminosity. Possible energy injection, e.g., from fallback accretion \citep{rosswog2007} or Blandford-Payne mechanism \citep{ma2018} could also enhance the brightness of the kilonova. For these cases, future follow-up observations of NSBH merger events would be able to find a bit more associated kilonovae. However, model diversity and possible energy injection would not affect the main conclusion of this work. 

Although most NSBH mergers are predicted to be plunging events, since the NSs are usually charged, some detectable EM signals, i.e., fast radio bursts or short-duration X-ray bursts, can be produced during the final merger phase for NSBH binaries \citep{zhang2019,dai2019,sridhar2021}. Moreover, some CBC events are believed to be embedded in the accretion disks of active galactic nuclei \citep[AGNs; e.g.,][]{cheng1999,mckernan2020}. The disrupted NSBH mergers in AGN disks can power sGRBs and kilonova emissions. Due to the dense atmosphere of the AGN disks, most of sGRB jets would be choked and hence it is difficult for us to observe their gamma-ray emission \citep{perna2021,zhu2021a}. The cocoon cooling emission and kilonova emission would be outshone by the disk emission. However, one can possibly detect the cocoon shock breakout signals in the soft X-ray band and ejecta shock breakout signals in the ultraviolet band. There might be also associated high-energy neutrino emission from choked gamma-ray jets \citep{zhu2021b} and subsequent two shock breakouts \citep{zhu2021c}, potentially detectable by IceCube. NSBH mergers in AGN disks are potential sources for future joint EM, neutrino, and GW multi-messenger observations.

\acknowledgments

We thank Ying Qin for valuable comments. The work of J.P.Z is partially supported by the National Science Foundation of China under Grant No. 11721303 and the National Basic Research Program of China under grant No. 2014CB845800. Y.W.Y is supported by the National Natural Science Foundation of China under Grant No. 11822302, 11833003. Y.P.Y is supported by National Natural Science Foundation of China grant No. 12003028 and Yunnan University grant No.C176220100087. H.G. is supported by the National Natural Science Foundation of China under Grant No. 11690024, 12021003, 11633001. Z.J.C is supported by the National Natural Science Foundation of China (No. 11690023). L.D.L. is supported by the National Postdoctoral Program for Innovative Talents (grant No. BX20190044), China Postdoctoral Science Foundation (grant No. 2019M660515), and “LiYun” postdoctoral fellow of Beijing Normal University.

\bibliography{NSBH}{}
\bibliographystyle{aasjournal}

\end{document}